\documentclass[10pt, twocolumn]{revtex4}
\usepackage{epsf}
\usepackage{amsmath}
\usepackage{amssymb}
\usepackage{epsfig}
\usepackage{latexsym}
\usepackage{amsfonts}
\usepackage{graphicx}%
\usepackage{varioref}
\usepackage{ifthen}
\begin{document}
\parindent 0mm
\setlength{\parskip}{\baselineskip}
\thispagestyle{empty}
\pagenumbering{arabic}
\setcounter{page}{1}
\mbox{ }

\preprint{UCT-TP-271/08}
\preprint{MZ-TH/08-***}

\title{Up and down quark  masses from Finite Energy QCD sum rules to five loops}

\author{C. A. Dominguez}
\affiliation{Centre for Theoretical Physics and Astrophysics, University of
Cape Town, Rondebosch 7700, South Africa, and \\ Department of Physics, Stellenbosch University, Stellenbosch 7600, South Africa}
\author{N. F. Nasrallah}
\affiliation{Faculty of Science, Lebanese University, Tripoli, Lebanon}
\author{R. H. R\"ontsch}
\affiliation{Centre for Theoretical Physics and Astrophysics, University of
Cape Town, Rondebosch 7700, South Africa}
\author{K. Schilcher}
\affiliation{Institut f\"{u}r Physik, Johannes Gutenberg-Universit\"{a}t,
Staudingerweg 7, D-55099 Mainz, Germany}
\date{\today}
\begin{abstract}
The up and down  quark masses are determined from an optimized QCD Finite Energy Sum Rule (FESR) involving the correlator of axial-vector divergences, to five loop order in Perturbative QCD (PQCD), and including leading non-perturbative QCD and higher order quark mass corrections. This FESR is designed to reduce considerably the systematic uncertainties arising from the (unmeasured) hadronic resonance sector, which in this framework contributes less than 3-4\% to the quark mass. This is achieved by introducing an integration kernel in the form of a second degree polynomial, restricted to vanish at the peak of the two lowest lying resonances. The driving hadronic contribution is then the pion pole, with parameters well known from experiment. The determination is done in the framework of Contour Improved Perturbation Theory (CIPT), which exhibits a very good convergence, leading to a remarkably stable result in the unusually wide window $s_0 = 1.0 - 4.0 \;\mbox{GeV}^2$, where $s_0$ is the radius of the integration contour in the complex energy (squared) plane. The results are: $m_u(Q= 2 \;\mbox{GeV}) = 2.9 \pm 0.2 $ MeV, $m_d(Q= 2 \;\mbox{GeV}) = 5.3 \pm 0.4$ MeV, and $(m_u + m_d)/2 = 4.1 \pm 0.2$ MeV (at a scale $Q=2$ GeV).\\
\end{abstract}
\maketitle
\noindent
The up and down quark masses are very important  parameters of the QCD sector of the Standard Model. They measure the strength of chiral $SU(2) \otimes SU(2)$ and flavour $SU(2)$ symmetry breaking, and are believed to be responsible for the proton-neutron mass difference. For this reason, many attempts have been made in the past to determine their values
in various frameworks, e.g. QCD sum rules \cite{OLD1}-\cite{OLD2}, and Lattice QCD \cite{LATTICE}, while chiral perturbation theory provides information on quark mass ratios \cite{RATIO}. In the framework of QCD sum rules, the ideal correlators are those involving the light quark masses as overall multiplicative factors, e.g. pseudoscalar correlators. Unfortunately, the corresponding hadronic resonance spectral functions are not realistically measurable in these channels (e.g. from $\tau$ decays). Channels where they are known from data (vector, axial-vector) involve the quark masses as subleading terms, and thus complicate their determination. The calculation of the scalar (pseudoscalar) correlator in Perturbative QCD (PQCD) has been improving over the years \cite{PQCD4}, reaching recently the five-loop order \cite{PQCD5}. Such level of precison is, sadly, not matched in the hadronic sector. In fact, the lack of direct experimental  data on the resonance spectral functions results in a serious systematic uncertainty, which  limits the accuracy of quark mass determinations in this framework. While in both the non-strange and the strange pseudoscalar channels there are a couple of resonances with known masses and widths \cite{PDG}, this is not enough to fully reconstruct the spectral functions. In fact,  inelasticity and non-resonant background are realistically impossible to model. Given this situation, a possible way of improving the precision of quark mass determinations is to reduce considerably the numerical contribution of hadronic resonances. We have recently achieved this in the strange pseudoscalar channel \cite{COND}-\cite{ms} by using new Finite Energy Sum Rules (FESR) involving an integration kernel in the form of a second degree polynomial which vanishes at the peaks of the two known resonances. This resonance quenching has the effect of reducing considerably (by up to one order of magnitude) the impact of this term, so that the pseudoscalar meson pole becomes the dominant hadronic contribution. The strange quark-mass \cite{ms} thus determined is essentially free of systematic uncertainties, and subject to future improvement from more precise determinations of $\Lambda_{QCD}$ and the gluon condensate, as well as from higher order  PQCD calculations (six loop and beyond). In addition,  the strange quark mass at a given fixed scale, e.g. $m_s(2 \mbox{GeV})$ turns out to be remarkably stable in a very wide range of values of $s_0$, the upper limit of integration in the FESR. In fact, this range is $s_0 = 1 - 4\; \mbox{GeV}^2$, a rather unusually broad window. In contrast, without the use of the new integration kernel, this window shrinks considerably to $s_0 = 1 - 2 \;\mbox{GeV}^2$, and  the ghost of the uncontrolable systematic uncertainties is resurrected.\\
In this paper we use this FESR in the light  pseudoscalar channel to find the value of the up and down quark masses to five-loop order in PQCD, and including the leading vacuum condensates and higher order quark mass corrections. We use the framework of  Contour Improved Perturbation Theory (CIPT) \cite{CIPT}, which has proven far more stable than Fixed Order Perturbation Theory (FOPT)  (see \cite{ms}), at least in the type of FESR we are using, with similar conclusions drawn from  a precision determination of $\alpha_s(M_\tau)$ from $\tau$ - decay in \cite{ALEPH2}.
We begin by introducing the correlator of  axial-vector divergences
\begin{equation}
\psi_{5} (q^{2}) = i\int d^{4}  x \; e^{i q x}
<|T(\partial^\mu A_{\mu}(x), \partial^\nu A_{\nu}^{\dagger}(0))|>,
\end{equation}
where $\partial^\mu A_{\mu}(x) = (m_u + m_d) :\overline{d}(x) \,i \, \gamma_{5}\, u(x):\;$ is the divergence of the  axial-vector current. To simplify the notation  we shall use in the sequel $m_u + m_d \equiv m$. Finite Energy QCD Sum Rules  involving the second derivative of this correlator follow from Cauchy's theorem in the complex energy-squared, s - plane, Fig. 1 (for details on the derivation see \cite{ms}) , i.e.
\begin{equation}
\begin{split}
 &-\frac{1}{2\pi i}\oint_{C(|s_0|)}ds \;\psi_{5}^{''}(s)|_{QCD}[F(s) - F(s_0)] =   \\  &2f_\pi^2 M_\pi^4 \Delta_5(M_\pi^2)
+
\frac{1}{\pi}  \int_{s_{th}}^{s_0}
ds Im \psi_{5}(s)|_{RES}\Delta_5(s)  ,
\end{split}
\end{equation}
where $s_{th} = (3 M_\pi)^2$ is the resonance threshold, $f_\pi = 92.4 \pm 0.3 \;\mbox{MeV}$, the function $F(s)$ is
\begin{eqnarray}
F(s) = &-& s \left(s_0 - a_0\,\frac{s_0^2}{2} - a_1\, \frac{s_0^3}{3} \right) + \frac{s^2}{2} \nonumber \\ &-& a_0\, \frac{s^3}{6} - a_1\, \frac{s^4}{12} \;,
\end{eqnarray}
and the integration kernel $\Delta_5(s)$ is
\begin{equation}
\Delta_5(s) = 1 - a_0 \;s - a_1\; s^2 \;,
\end{equation}
where $a_{0}$, and $a_1$ are free parameters to be fixed by the requirement that $\Delta_5(M_1^2) = \Delta_5(M_2^2) = 0$, with $M_{1,2}$ the masses of the two resonances in the  pseudoscalar channel, $\pi$(1460) and $\pi$(1830) \cite{PDG}. This gives
\begin{equation}
a_0 = 0.897  \; \mbox{GeV}^{-2} \; \;  a_1 = - 0.1806 \;\; \mbox{GeV}^{-4} \;.
\end{equation}
The function $\psi_{5}{''}(Q^2)$ in PQCD is given by
\begin{eqnarray}
&&\psi_5^{'' PQCD}(Q^2) = \frac{3}{8 \pi^2}\; \frac {\overline{m}^2(Q^2)}{ Q^2}\;  \left\{ 1 +\; \frac{11}{3}\; \frac{\alpha_s(Q^2)}{\pi}\; \right. \nonumber \\ &+& \left. (\frac{\alpha_s(Q^2)}{\pi})^2 \left[ -  \frac{35}{2} \zeta(3)
+  \frac{5071}{144}  \right] \nonumber \right. \\  &+& \left. O (\alpha_s^3) + O (\alpha_s^4) \phantom{\frac{1}{1}}\right\} ,
\end{eqnarray}
with $Q^2 \equiv - q^2$, and Renormalization Group (RG) improvement has been used to dispose of the logarithmic terms.
The rather long four- and five-loop expressions can be found in \cite{PQCD4}-\cite{PQCD5}. Since it will turn out that $s_0 \simeq 1 - 4 \;\mbox{GeV}^2$, we have taken $n_F =3$ in Eq.(6) and in the sequel. Using two flavours, instead, changes the results by a tiny amount well within the final error bars.
\begin{figure}
[ht]
\begin{center}
\includegraphics[height=1.9in, width=1.8in]
{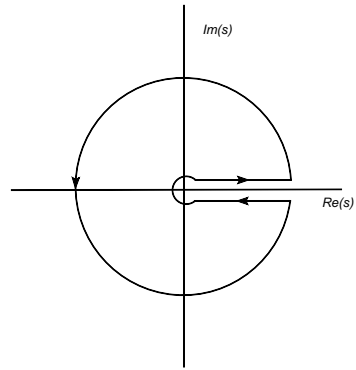}
\caption{Integration contour in the complex s-plane.}
\end{center}
\end{figure}
Contour Improved Perturbation Theory \cite{CIPT} has been shown to provide better convergence than FOPT in the QCD analysis of the vector and axial-vector correlators in tau-lepton decay \cite{CIPT}-\cite{CADTAU}. We found this to be also the case in our determination of $m_s$ \cite{ms}, and it will be confirmed  here for $m_{u,d}$ . Unlike the case of FOPT, where $\alpha_s(s_0)$ is frozen in Cauchy's contour integral and the (RG) is implemented after integration, in CIPT $\alpha_s$ is running and the RG is used before integrating. This is done  through a single-step numerical contour integration and using as input the strong coupling obtained  by solving numerically the RG Equation (RGE) for $\alpha_s(-s)$ . This technique achieves a partial resummation of the higher order logarithmic integrals, and improves the convergence of the PQCD series. In the case of the pseudoscalar correlator involving the running quark mass as an overall multiplicative factor, implementation of CIPT requires that not only the running coupling but also the running quark mass be integrated around the Cauchy contour. The running quark mass can be computed at each step by solving numerically the  corresponding RGE. We define the left hand side of Eq.(2) as
\begin{equation}
\delta_5(s_0)|_{QCD} \equiv - \frac{1}{2\pi i}
\oint_{C(|s_0|)}
ds \psi_{5}^{''}(s)|_{QCD}[F(s) - F(s_0)] ,
\end{equation}
and obtain in PQCD (for more details see \cite{ms})
\begin{widetext}
\begin{equation}
\delta_5(s_0)|_{PQCD} = \frac{\overline{m}^2(s_0)}{16 \pi^2} \,\sum_{j=0}^4 K_j \;
\frac{1}{2 \pi} \;
  \int_{-\pi}^{\pi} dx \;  \Big[ F(x) - F(s_0) \Big]
  [a_s(x)]^j \;
exp \Bigg[ - 2 i \sum_{M=0} \gamma_M \int_0^x\; dx' \;[a_s(x')]^{M+1} \Bigg]  \;,
\end{equation}
\end{widetext}
where the RGE for the mass and coupling have been used, $a_s(x) \equiv \alpha_s(x)/\pi$, and where
\begin{equation}
F(x) = \sum_{N=1}^4 (-)^N \; b_N \; s_0^N \; e^{iNx} \;.
\end{equation}
The constants above are ($n_F = 3$): $K_0= C_{01}$, $K_1 = C_{11} + 2 C_{12}$, $K_2 = C_{21} + 2 C_{22}$, $K_3 = C_{31} + 2 C_{32}$, $K_4 = C_{41} + 2 C_{42}$, with
$C_{01} = 6$, $C_{11} = 34$, $C_{12} = - 6$, $C_{21} = - 105 \;\zeta(3) + 9631/24$,
$C_{22} = - 95$, $C_{23} = 17/2$, $C_{31}= 4748953/864 - \pi^4/6 - 91519\; \zeta(3)/36 + 715 \;\zeta(5)/2$,
$C_{32} = - 6 \;[4781/18 - 475 \;\zeta(3)/8]$, $C_{33} = 229$, $C_{34} = - 221/16$, $C_{41} = 33 532.26$, $C_{42} = - 15 230. 6451$, $C_{43} = 3962.45493$, $C_{44} = - 534.052083$, $C_{45} = 24.1718750$, and $\zeta(x)$ is Riemann's zeta function.
Finally, $b_1= -(s_0 - a_0 s_0^2/2 - a_1 s_0^3/3)$, $b_2 = 1/2$, $b_3 = - a_0/6$, and $b_4 = - a_1/12$.
Regarding the value of $\Lambda_{QCD}$ entering $\alpha_s(s_0)$,  it can be extracted from the strong coupling obtained from $\tau$-decay \cite{PDG}, \cite{DAVIER}. We use the latest high precision result \cite{ALEPH2} leading to $\Lambda_{QCD} = 365 - 397$ MeV.\\
\begin{figure}
\begin{center}
\includegraphics[height=1.8 in, width=1.8 in]
{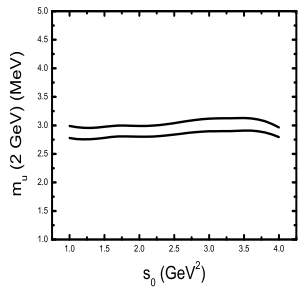}
\caption{Up quark mass at 2 GeV as a function of $s_0$ for $\Lambda_{QCD} = 365 \;(397)\; \mbox{MeV}$, upper (lower) curve, respectively.}
\end{center}
\end{figure}
\begin{figure}
\begin{center}
\includegraphics[height=1.8 in, width=1.8 in]
{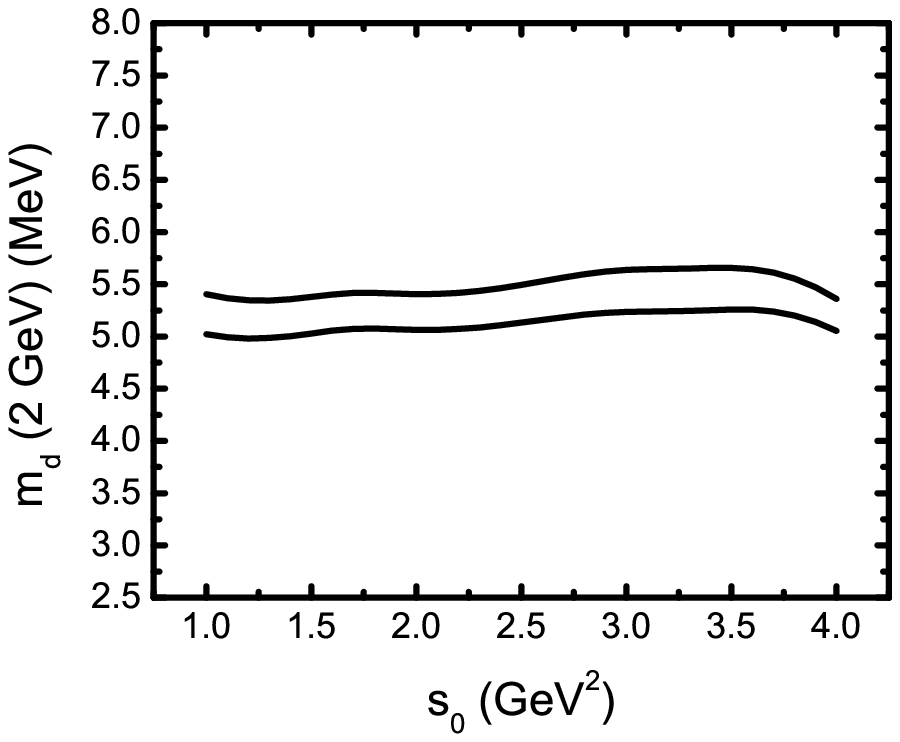}
\caption{Down quark mass at 2 GeV as a function of $s_0$ for $\Lambda_{QCD} = 365 \;(397)\; \mbox{MeV}$, upper (lower) curve, respectively.}
\end{center}
\end{figure}The contribution of the gluon condensate to the left hand side of Eq.(7) is
\begin{eqnarray}
&&\delta_5(s_0)|_{<G^2>} = \frac{1}{4} \;\;\frac{\overline{m}^2(s_0)}{s_0^2} \; \;\langle \frac{\alpha_s}{\pi} G^2\rangle|_{\mu_0}\;\;
\nonumber \\[0.3cm] &\times& \frac{1}{2\pi} \int_{-\pi}^\pi dx \;e^{-2 i x}
 \Big[F(x) - F(s_0)\Big] \nonumber \\ [0.3cm] &\times& \Big[ 1 + \frac{16}{9} a_s(\mu_0)+ \frac{121}{18} a_s(x)\Big]
 \nonumber \\ [0.3cm] &\times& exp  \Bigg[ - 2 i \sum_{M=0} \gamma_M  \int_0^x dx' [a_s(x')]^{M+1} \Bigg] \;,
\end{eqnarray}
where the scale $\mu_0 \simeq 1 \; \mbox{GeV}^2$ appears in connection with the removal of logarithmic quark mass singularities (see \cite{OLD2}). In obtaining this expression use was made of the result \cite{OLD2}
\begin{eqnarray}
&&\psi_5^{'' PQCD}(Q^2)|_{\langle G^2 \rangle} =\frac{m^2(Q)}{Q^6} \, \frac{1}{4}\, \langle \frac{\alpha_s}{\pi} G^2 \rangle|_{\mu_0}  \nonumber\\ [.3cm]
&\times& \left(1 +  \frac{16 \alpha_s(\mu_0)}{9 \pi}+ \frac{121\alpha_s(Q)}{18 \pi}\right) \;.
\end{eqnarray}
The quark condensate, the dimension six four-quark condensate, and the higher order quark mass corrections turn out to be negligible in this application. Turning to the hadronic sector,  we follow the procedure outlined in \cite{CADZEIT}, where the resonance part of the spectral function is written as a linear combination of two Breit-Wigner forms normalized at threshold according to chiral perturbation theory, i.e.
\begin{equation}
\frac{1}{\pi} Im \,\psi_5(s)|_{RES} = \frac{1}{3 (4 \pi)^4} \; \frac{M_\pi^4}{f_\pi^2} \; s \; BW(s) \;,
\end{equation}
where $BW(s)$ is a linear combination of two Breit-Wigner forms normalized to $BW(0) = 1$. The phase space factor above is shown in the chiral limit for simplicity. The full expression including the threshold at $s = 9 M_\pi^2$ is given in \cite{CADZEIT}.
Substituting Eqs.(8) and (10) in the sum rule Eq.(2), using  $\langle \frac{\alpha_s}{\pi} G^2\rangle|_{\mu_0} = 0.06\; \mbox{GeV}^4$ \cite{CADTAU}, $\Lambda_{QCD} = 365 - 397$ MeV \cite{ALEPH2}, and $m_u/m_d = 0.553$ from CPT \cite{RATIO}, we find the results shown in Fig. 2 and Fig. 3, for $m_u(2 \;\mbox{GeV})$ and $m_d(2 \;\mbox{GeV})$, respectively. The stability of the results is exceptional, and the duality window unusually broad. In order to achieve a reasonable, but conservative error, we have analyzed the impact of the various contributions to the quark masses. Some of these contributions have opposite effects, e.g. decreasing the hadronic resonance term has the same effect as increasing the QCD contribution, and vice-versa. Allowing for a 30\% uncertainty (upwards or downwards) in the resonance spectral function, multiplying (dividing) the gluon condensate by a factor of two, and assuming the unknown six-loop PQCD term to be equal to the five-loop one, and taking into account the current uncertainty in $\Lambda_{QCD}$, we arrive at the following results
\begin{eqnarray}
m_u (2 \;\mbox{GeV}) = 2.9 \; \pm \; 0.2 \; \mbox{MeV} \;, \nonumber \\ [.3cm]
m_d (2 \;\mbox{GeV}) = 5.3 \; \pm \; 0.4 \; \mbox{MeV} \; \nonumber \\ [.3cm]
m_{ud} \equiv \frac{m_u + m_d}{2} = 4.1 \; \pm \; 0.2 \; \mbox{MeV}
\end{eqnarray}
The allowance of a $\pm 30 \%$ uncertainty in the hadronic resonance contribution impacts the quark masses at the level of only 1 \%.
Our choice of integration kernel, Eq.(4), is by no means unique. We have tried several other functional forms with the same constraint, in particular the so called {\it pinched} kernels \cite{PINCH}. These kernels are problematic in this application, as the relative contribution of the poorly known spectral function is much larger, and the stability region is considerably reduced. The same effect was observed in the case of the strange quark mass[9]. We have also checked that using polynomials of higher order, again restricted to vanish at the resonance peaks, has a negative impact on the QCD sector of the FESR due to the emergence of unknown higher dimensional vacuum condensates, and
the divergence of polynomial coefficients. From the point of view of
errors our polynomial parametrization, Eq. (4), represents the optimal choice.
Disregarding the stability region as a criterion, results for the quark masses with other choices for the kernel are consistent within errors with the results above. In particular, for a given hadronic resonance spectral function, e.g. Eq.(12), if we remove the kernel altogether ($\Delta_5(s) \equiv 1$) the stability region shrinks to the narrow window $s_0 = 1 - 2\;
\mbox{GeV}^2$,  and the systematic uncertainties, from the (unmeasured) hadronic resonance spectral function, make a comeback.\\
In comparing these results with others from QCD sum rules, one should exercise extreme care, as older determinations used much lower values of $\Lambda_{QCD}$, and less terms in PQCD (with the exception of the first five-loop result of \cite{PQCD5}). However, the most important difference with previous determinations lies in our choice of integration kernel, which reduces considerably systematic uncertainties from the resonance sector. The results above compare  favourably with some recent lattice QCD determinations, e.g.
for the average:  $ m_{ud} = 3.85 \; \pm \; 0.42 \;\mbox{MeV}$ \cite{LATTICE2}, $ m_{ud} = 3.72 \; \pm \; 0.41 \;\mbox{MeV}$ \cite{LATTICE3}, and $ m_{ud} = 3.3 \; \pm \; 0.2 \;\mbox{MeV}$ \cite{LATTICE4}, where all errors have been combined in quadrature. However, one other lattice QCD result of $ m_{ud} = 2.527 \; \pm \; 0.047 \;\mbox{MeV}$  \cite{LATTICE5} is much smaller than these.\\
This work was supported in part by  NRF (South Africa) and DFG (Germany)
 \textsl{}

\end{document}